\documentclass[11pt]{article}

\usepackage[preprint]{acl}

\usepackage{times}
\usepackage{latexsym}
\usepackage{amsmath}
\usepackage[T1]{fontenc}
\usepackage[utf8]{inputenc}
\usepackage{xcolor}
\usepackage{microtype}
\usepackage{inconsolata}
\usepackage{graphicx}
\usepackage{tikz}
\usepackage{url}
\usetikzlibrary{shapes.geometric, arrows.meta, positioning}
\usepackage{multirow}

\title{
Do Better Volatility Forecasts Lead to Better Portfolios?\\
Evidence from Graph Neural Networks
\thanks{Code available at \url{https://github.com/waderylan/sp500-gnn}.}
}

\author{
\textbf{Rylan Wade}\\[6pt]
University of Southern California
}
\begin{document}
\maketitle

\begin{abstract}
This paper tests whether graph neural networks improve realized volatility forecasts and whether those forecasts improve portfolio performance. Using weekly realized volatility for 465 S\&P 500 equities from 2015--2025, Heterogeneous Autoregressive and Long Short-Term Memory baselines are compared against GraphSAGE models built on rolling correlation, sector, and Granger-causal graphs, with and without macro regime features. The empirical finding is that the model with the lowest forecast MSE, the model with the highest cross-sectional ranking accuracy, and the model with the highest portfolio Sharpe ratio are three different models. Forecast accuracy, ranking quality, and portfolio performance are related but not interchangeable objectives. Graph volatility models add value only when the portfolio rule can exploit the cross-sectional structure they encode.
\end{abstract}

\section{Introduction}
Volatility forecasting is central to portfolio construction, risk management, and position sizing. Yet volatility does not move in isolation. When turbulence hits one stock, it often spreads through sector peers, correlated positions, and broader market relationships. Traditional models capture persistence in each stock's own volatility history, but they do not directly model how volatility propagates across the market. 

In contrast, Graph Neural Networks (GNNs) offer a natural way to model these relationships. By representing stocks as nodes and relationships as edges, GNNs can propagate volatility signals across the market in a way univariate models cannot directly capture, conditioning each stock's forecast on the recent behavior of its neighbors. In principle, this should help when volatility shocks travel through correlated names or economically linked firms. In practice, however, graph structure is only useful if the edges encode relationships that remain informative out of sample. A poorly specified graph may add noise, propagate stale information, or compress meaningful differences across stocks.

This paper asks two related questions. First, do GNNs produce better weekly realized volatility forecasts than strong classical and deep learning baselines? Second, do those forecast improvements translate into better portfolio performance? The distinction matters because lower forecast error does not automatically imply higher investment value. A model can improve mean squared error while producing forecasts that are poorly calibrated, weakly ranked across stocks, or poorly suited to a given portfolio construction.

We run a controlled comparison across three graph construction strategies: dynamic return correlations, GICS sector membership, and Granger causality. All feed into an identical GraphSAGE backbone, so that performance differences reflect the graph construction rather than changes in model architecture. The Heterogeneous Autoregressive model (HAR) and the Long Short-Term Memory network (LSTM) serve as the primary baselines, representing classical and deep learning approaches to univariate volatility forecasting. A macro feature set capturing VIX, credit spreads, and cross-sectional correlation structure is then layered in to isolate the contribution of regime conditioning from graph structure. Finally, each model is evaluated under four portfolio constructions: minimum variance, inverse-volatility weighting, volatility targeting, and long-short. 

Taken together, this design tests whether graph-based volatility forecasts improve accuracy and portfolio performance, and under what modeling and portfolio conditions those improvements matter.

\section{Data and Features}

\subsection{Data Universe and Target}

We focus on S\&P 500 equities because the universe is large enough to make cross-sectional graph structure meaningful while remaining liquid enough for realistic portfolio simulation. The dataset consists of daily OHLCV data for S\&P 500 equities from January 2015 through December 2025, sourced via yfinance. After filtering to stocks with at least 95\% price coverage, the final universe contains 465 stocks. This is a current-constituent universe and therefore subject to survivorship bias, since firms that left the index before the end of the sample are excluded.

The prediction target is weekly realized volatility, computed as the annualized standard deviation of daily log returns within each calendar week, excluding weeks with fewer than three trading days. Weekly aggregation reduces microstructure noise and reflects realistic rebalancing frequency at this universe size.

\subsection{Feature Groups}

The feature set is designed to capture volatility information at three levels: the stock's own history, its recent trading behavior, and the broader market regime.

Volatility features include realized volatility over 5, 10, 21, and 63-day lookback windows, along with a short-to-long ratio capturing spikes relative to the longer-run baseline \cite{andersen1998}. These features capture volatility persistence across short, medium, and long horizons.

Return and volume features include 5- and 20-day price momentum, and log-transformed rolling volume at both horizons with a short-to-long volume ratio. These features capture recent price dislocations and abnormal trading activity that may precede changes in volatility.

Macro and regime features include VIX level and change, SPY volatility and returns, the 10-year to 2-year Treasury spread, investment-grade credit spreads, average pairwise correlation, and graph density. These variables provide explicit regime awareness beyond each stock’s own history.

\subsection{Preprocessing and Splits}

Stock-level features are winsorized cross-sectionally at the 1st and 99th percentiles each week and then z-scored. This limits the influence of extreme observations while preserving relative cross-sectional differences. Macro features are normalized using training-period statistics rather than cross-sectional scaling, since they are market-level variables shared across all stocks in a given week.

The data is split chronologically, with 2015--2022 used for training, 2023 for validation, and 2024--2025 for testing. All features are lagged so that forecasts for week \(t+1\) use only information available through the end of week \(t\).

\section{Models}

We evaluate a progression of models from simple to complex to isolate the contribution of each component in volatility forecasting. The comparison begins with the Heterogeneous Autoregressive (HAR) model as a strong linear baseline, followed by a Long Short-Term Memory (LSTM) network to capture nonlinear temporal dynamics at the individual stock level. We then introduce Graph Neural Networks (GNNs) to incorporate cross-sectional dependencies between stocks, using a shared GraphSAGE backbone across multiple graph constructions. This design allows us to attribute performance differences to specific additions: sequence modeling capacity in the LSTM, cross-stock information in the GNN, graph topology across the three graph variants, and regime conditioning in the macro feature set.

Table~\ref{tab:hparam_selection} summarizes the main fixed and tuned parameters by model component. HAR and LSTM settings were fixed as benchmark specifications, while the GNN architecture, correlation threshold, and rolling-window choices were selected through validation grid search on 2023 data before final evaluation on the 2024--2025 test set.

\begin{table*}[t]
\centering
\caption{Hyperparameter and lookback search space for the forecasting models.}
\label{tab:hparam_selection}
\small
\begin{tabular}{lllc}
\hline
\textbf{Component} & \textbf{Parameter} & \textbf{Values considered} & \textbf{Selected value} \\
\hline
HAR & Volatility lags & 5, 21, 63 trading days & 5, 21, 63 \\
LSTM & Architecture & fixed baseline & 2 layers, hidden 64, dropout 0.3 \\
\hline
GNN-Correlation & Correlation threshold $\theta$ & 0.3, 0.5, 0.7 & 0.3 \\
GNN-Correlation + Macro & Rolling correlation window & 21, 63, 126, 252 trading days & Evaluated as separate variants \\
GNN-Granger & Granger lag & 5 trading days & 5 \\
\hline
GNN architecture & Learning rate & $10^{-4}$, $3\cdot10^{-4}$, $10^{-3}$, $3\cdot10^{-3}$ & $10^{-3}$ \\
GNN architecture & Hidden dimension & 64, 128, 256 & 256 \\
GNN architecture & GraphSAGE layers & 2, 3 & 3 \\
GNN architecture & Dropout & 0.1, 0.3 & 0.3 \\
GNN architecture & Normalization & none, BatchNorm, GraphNorm & none \\
\hline
\end{tabular}
\end{table*}

\subsection{HAR (Heterogeneous Autoregressive Model)}

The Heterogeneous Autoregressive model (HAR) \cite{corsi2009} is a standard linear benchmark for realized volatility forecasting. It predicts next-period volatility as a linear function of average realized volatility over three lookback windows: 5, 21, and 63 trading days, capturing volatility's well-documented persistence across short, medium, and long horizons. Two variants are included. Per-stock HAR fits an independent regression for each of the 465 stocks using only that stock's own history. Pooled HAR fits a single set of coefficients across the entire universe simultaneously. HAR features are computed directly from the raw volatility series rather than the normalized feature tensor, keeping it a pure autoregression.

\subsection{LSTM}

We implement a two-layer Long Short-Term Memory network \cite{hochreiter1997} with hidden size 64 and dropout of 0.3, followed by a linear output layer. It receives a four-week rolling window of all ten stock-level features and processes each stock independently, with no information passing between stocks. Its role in the model comparison is specific: to isolate the contribution of deep learning capacity from the contribution of graph structure. This makes the LSTM a useful benchmark for separating stock-level sequence modeling from the additional cross-stock information introduced by the GNN variants.

\subsection{GNN Backbone}

All three GNN variants share an identical GraphSAGE backbone \cite{hamilton2017}. GraphSAGE operates via neighborhood aggregation: for each stock node, it computes the mean of neighboring stocks' feature vectors, concatenates this with the node's own features, and passes the result through a learned linear transformation. GNN hyperparameters, including learning rate, hidden dimension, dropout, number of layers, correlation threshold, and graph-specific lookback windows, were selected via grid search on the 2023 validation set. HAR and LSTM were kept as fixed benchmark specifications.

\begin{table*}[t]
\centering
\caption{Rolling correlation graph statistics for three representative weeks under the $|\rho| \geq 0.30$ edge rule, where $\rho$ denotes rolling Pearson return correlation. The universe has 465 stocks and 107,880 possible undirected edges. Avg.\ $|\rho|$ is computed across all stock pairs, not only connected pairs.}
\label{tab:graph_density_examples}
\small
\begin{tabular}{llrrrr}
\hline
\textbf{Regime} & \textbf{Week} & \textbf{Edges} & \textbf{Density} & \textbf{Mean Deg.} & \textbf{Avg. $|\rho|$} \\
\hline
Sparse week & 2018-01-22 & 9,876 & 0.092 & 42.5 & 0.147 \\
COVID shock & 2020-03-23 & 100,609 & 0.933 & 432.7 & 0.547 \\
Calm test week & 2024-12-09 & 18,775 & 0.174 & 80.8 & 0.187 \\
\hline
\end{tabular}
\end{table*}

\subsection{Graph Construction Strategies}

The three graph constructions encode different views of how stocks are related. GNN-Correlation tracks live volatility co-movement, GNN-Sector encodes fixed industry classification, and GNN-Granger captures lagged causal relationships. They differ in whether they reflect current market structure and whether they update over time, two properties that turn out to matter for both forecast accuracy and portfolio performance.

\textbf{GNN-Correlation} connects two stocks with an undirected edge if their rolling Pearson return correlation satisfies $|\rho| \geq 0.30$. The no-macro correlation GNN uses a 252-day rolling correlation window, while macro-conditioned correlation GNNs are reported separately for 21-, 63-, 126-, and 252-day windows. The graph is recomputed each week as the rolling window advances, making it the most dynamic of the three constructions. During calm regimes the graph is moderately dense; during crisis periods it approaches full connectivity as stocks move together. Correlation graphs for all weeks were precomputed and cached to disk, reducing per-run overhead to under one second.

Table~\ref{tab:graph_density_examples} shows how much the rolling correlation graph changes across regimes under the same $|\rho| \geq 0.30$ threshold. In a sparse 2018 week, only 9.2\% of possible stock pairs crossed the threshold, giving the average stock about 43 neighbors. During the COVID shock, the same rule produced an almost complete graph, with density 0.933 and average degree 432.7 out of 464 possible neighbors. A calmer test-period week in December 2024 falls between these extremes, with density 0.174 and average degree 80.8. This variation matters because density changes what the GNN is aggregating. In sparse weeks, the graph passes information through selected co-movement relationships. In crisis weeks, the graph becomes broad enough that aggregation starts to look market-wide.

\textbf{GNN-Sector} connects stocks sharing the same GICS sector, updated annually to reflect historical reclassifications. It encodes fundamental economic relationships rather than return co-movement, and changes only once per year, making it the most stable of the three constructions.

\textbf{GNN-Granger} is a directed graph. An edge runs from stock A to stock B when A's past returns contain statistically significant predictive information about B's returns beyond B's own history \cite{granger1969}. Tests were run for all ordered pairs using a five-day lag, with Bonferroni correction applied to control for multiple comparisons, yielding 13,886 directed edges. Unlike the other two graphs it is static, computed once over the training period, and its directionality is preserved through message passing rather than symmetrized.
 
\subsection{GNN Ensemble}

The ensemble combines predictions from all three GNN variants by taking a weighted average at inference time, with each model's weight proportional to the inverse of its validation MSE. Its role is to test whether the three graph constructions capture complementary cross-sectional information that no single graph encodes on its own.

\section{Evaluation Framework}

Each model is evaluated in three ways. Forecast accuracy measures how close the predicted weekly volatility is to realized volatility. Cross-sectional ranking measures whether the model correctly orders stocks from low to high future volatility. Portfolio simulation then tests whether those forecasts are useful once they are turned into actual positions.

This separation is important because each layer can fail differently. A model can have strong MSE but weak rankings, or rank volatility well while still producing poor portfolio returns if the construction uses the signal badly. For that reason, the results are interpreted across all three layers rather than treating forecast accuracy alone as the final outcome.

\subsection{Statistical Accuracy}

Mean Squared Error (MSE) is the primary training objective and model comparison metric. It penalizes large mispredictions more heavily than Mean Absolute Error (MAE), which is appropriate for volatility forecasting because failing to anticipate a spike carries disproportionate portfolio consequences. MAE, \(R^2\), and directional accuracy are reported as secondary metrics, capturing less outlier-sensitive error, scale-free explanatory power, and the ability to predict whether volatility rises or falls.

\subsection{Cross-Sectional Ranking}

Whether a model produces accurate point forecasts is a separate question from whether it correctly orders stocks by predicted volatility. Rank IC, the cross-sectional Spearman rank correlation between predicted and actual realized volatility across all stocks at each test week, measures this directly. The IC information ratio, defined as mean IC divided by its standard deviation across weeks, measures signal consistency and penalizes models that rank well occasionally but cannot be relied upon week to week.

\subsection{Portfolio Simulation}

Statistical metrics alone cannot answer whether better forecasts lead to better portfolios. Four portfolio constructions are evaluated over the 103 test weeks, each translating predicted volatility into position sizing in a different way.

Inverse-volatility weighting allocates capital to each stock in proportion to the inverse of its predicted volatility. Stocks forecast to be calm receive larger weights; stocks forecast to be turbulent receive smaller ones.

The long-short construction goes long the quintile of stocks with the lowest predicted volatility and shorts the quintile with the highest.

Volatility targeting scales overall portfolio exposure up or down each week to hit a fixed annualized volatility target, using predicted volatility to anticipate when to reduce exposure before turbulent periods arrive.

Minimum variance optimization uses predicted volatility as the diagonal of the covariance matrix and solves for the portfolio with the lowest expected variance subject to weight constraints.

Together, these constructions test whether forecast quality survives translation into portfolio decisions with different sensitivities to calibration, ranking, and regime timing.

An equal-weight portfolio serves as the passive baseline. All portfolios rebalance weekly with transaction costs applied at 10 basis points per unit of turnover, and Sharpe ratios are computed against the actual 3-month Treasury bill rate.

\section{Results}

The evaluation covers 103 test weeks from January 2024 through December 2025.

\subsection{Forecast Accuracy}

\begin{table*}[t]
\centering
\caption{Forecast accuracy on 103 test weeks from January 2024 through December 2025. Lower MSE and MAE are better; higher $R^2$ and directional accuracy are better.}
\label{tab:ml_metrics}
\small
\begin{tabular}{lcccc}
\hline
\textbf{Model} & \textbf{MSE} & \textbf{MAE} & \textbf{R$^2$} & \textbf{DA} \\
\hline
\multicolumn{5}{l}{\textit{Baselines}} \\
HAR per-stock          & 0.0329 & 0.1129 & 0.131 & 0.707 \\
HAR pooled             & 0.0331 & 0.1133 & 0.124 & 0.703 \\
LSTM                   & 0.0324 & 0.1096 & 0.142 & 0.709 \\
\hline
\multicolumn{5}{l}{\textit{Graph models without macro features}} \\
GNN-Correlation        & 0.0322 & 0.1076 & 0.148 & 0.712 \\
GNN-Sector             & 0.0336 & 0.1208 & 0.110 & 0.682 \\
GNN-Granger            & 0.0337 & 0.1190 & 0.108 & 0.688 \\
GNN-Ensemble           & 0.0320 & 0.1127 & 0.153 & 0.700 \\
\hline
\multicolumn{5}{l}{\textit{Graph models with macro features}} \\
GNN-Corr.\ + Macro 21d   & 0.0311 & 0.1043 & 0.177 & 0.725 \\
GNN-Corr.\ + Macro 63d   & 0.0298 & 0.1042 & 0.210 & 0.722 \\
GNN-Corr.\ + Macro 126d  & 0.0321 & 0.1080 & 0.150 & 0.719 \\
GNN-Corr.\ + Macro 252d  & 0.0309 & 0.1070 & 0.183 & 0.720 \\
GNN-Sector + Macro        & 0.0315 & 0.1107 & 0.166 & 0.704 \\
GNN-Granger + Macro       & 0.0314 & 0.1077 & 0.168 & 0.714 \\
GNN-Ensemble + Macro      & 0.0316 & 0.1073 & 0.164 & 0.715 \\
\hline
\end{tabular}
\end{table*}

Table~\ref{tab:ml_metrics} reports forecast accuracy across all models on the 103 test weeks. HAR per-stock achieves a test MSE of 0.0329 and is the primary baseline. HAR pooled is similar at 0.0331, confirming that sharing coefficients across 465 stocks costs little when volatility dynamics are broadly similar. The LSTM improves modestly to 0.0324, showing that sequence modeling capacity alone provides a limited edge over the linear benchmark.

Graph structure helps most when the graph is dynamic. Without macro features, GNN-Correlation and GNN-Ensemble reach MSE values of 0.0322 and 0.0320, outperforming both HAR variants and the LSTM. GNN-Sector and GNN-Granger move in the opposite direction, posting MSE values of 0.0336 and 0.0337. Static sector membership and Granger-causal edges do not improve point-forecast accuracy on their own. The gap between correlation and sector graphs indicates that dynamic return co-movement carries more timely predictive signal than fixed industry classification.

Macro-conditioned graph models produce the strongest point forecasts. GNN-Corr.\ + Macro 63d achieves the lowest MSE, 0.0298, with the highest $R^2$, 0.210. GNN-Corr.\ + Macro 21d has the highest directional accuracy, 0.725, but not the lowest MSE. This distinction matters because the best model depends on whether the objective is point accuracy, directional movement, ranking, or portfolio performance.

\subsection{Cross-Sectional Ranking}

\begin{table*}[t]
\centering
\caption{Cross-sectional ranking over 103 test weeks. Mean IC is the weekly Spearman rank correlation between predicted and realized volatility. ICIR is mean IC divided by its weekly standard deviation.}
\label{tab:rank_ic}
\small
\begin{tabular}{lcccc}
\hline
\textbf{Model} & \textbf{Mean IC} & \textbf{ICIR} & \textbf{Top-Q HR} & \textbf{Pair Acc.} \\
\hline
\multicolumn{5}{l}{\textit{Baselines}} \\
HAR per-stock          & 0.405 & 3.473 & 0.486 & 0.641 \\
HAR pooled             & 0.392 & 3.437 & 0.478 & 0.636 \\
LSTM                   & 0.429 & 4.363 & 0.499 & 0.648 \\
\hline
\multicolumn{5}{l}{\textit{Graph models without macro features}} \\
GNN-Correlation        & 0.417 & 3.440 & 0.492 & 0.645 \\
GNN-Sector             & 0.383 & 3.399 & 0.479 & 0.633 \\
GNN-Granger            & 0.375 & 3.663 & 0.473 & 0.629 \\
GNN-Ensemble           & 0.416 & 3.577 & 0.494 & 0.645 \\
\hline
\multicolumn{5}{l}{\textit{Graph models with macro features}} \\
GNN-Corr.\ + Macro 21d   & 0.412 & 3.915 & 0.496 & 0.643 \\
GNN-Corr.\ + Macro 63d   & 0.426 & 3.657 & 0.502 & 0.648 \\
GNN-Corr.\ + Macro 126d  & 0.415 & 3.525 & 0.490 & 0.644 \\
GNN-Corr.\ + Macro 252d  & 0.429 & 3.680 & 0.502 & 0.649 \\
GNN-Sector + Macro        & 0.428 & 3.850 & 0.497 & 0.647 \\
GNN-Granger + Macro       & 0.429 & 4.064 & 0.494 & 0.648 \\
GNN-Ensemble + Macro      & 0.438 & 3.935 & 0.501 & 0.653 \\
\hline
\end{tabular}
\end{table*}

Point-forecast MSE measures average loss across all stocks and weeks, but portfolio construction depends more on cross-sectional ranking: which stocks will be the calmest or most turbulent in a given week. Rank IC measures this directly, and the ranking results in Table~\ref{tab:rank_ic} tell a sharper story than MSE alone.

The strongest mean Rank IC is achieved by GNN-Ensemble + Macro at 0.438. GNN-Granger + Macro, GNN-Corr. + Macro 252d, GNN-Sector + Macro, and LSTM all cluster closely behind, with mean IC values between 0.428 and 0.429. LSTM remains especially competitive on ranking, with the highest ICIR in the table, 4.363. GNN-Sector and GNN-Granger are the weakest on both MSE and Rank IC without macro features, ruling out the possibility that their under-performance is an artifact of the loss function.

These results reinforce that graph connectivity is not generically useful; its value depends on whether the edges reflect current market structure. Dynamic return correlations appear to capture timely cross-sectional information, while sector labels and lagged Granger relationships are less effective without additional regime conditioning.

\subsection{The Role of Macro Features}

Macro features improve every graph model relative to its non-macro counterpart in forecast MSE. The largest point-forecast gain occurs for GNN-Corr.\ + Macro 63d, which reaches an MSE of 0.0298. Macro conditioning also substantially improves the weaker graph constructions: GNN-Sector improves from 0.0336 to 0.0315, and GNN-Granger improves from 0.0337 to 0.0314.

This pattern indicates that regime information and graph topology are complementary. When graph structure is weak or slow-moving, macro features provide market-wide context that helps recover forecast performance. The result is especially important for the later portfolio analysis, where macro-conditioned sector and Granger models produce the strongest minimum-variance results.

\subsection{Portfolio Performance}

\begin{table*}[t]
\centering
\caption{Minimum-variance portfolio performance over 103 test weeks with 10 bps transaction cost and max 5\% single-stock weight.}
\label{tab:port_minvar}
\small
\begin{tabular}{lccccc}
\hline
\textbf{Model} & \textbf{Ann.\ Ret.} & \textbf{Ann.\ Vol.} & \textbf{Sharpe} & \textbf{Max DD} & \textbf{Avg.\ Turn.} \\
\hline
\multicolumn{6}{l}{\textit{Baselines}} \\
HAR per-stock          & 0.113 & 0.099 & 0.635 & $-$0.127 & 1.012 \\
HAR pooled             & 0.124 & 0.102 & 0.729 & $-$0.119 & 0.913 \\
LSTM                   & 0.112 & 0.110 & 0.564 & $-$0.114 & 0.591 \\
\hline
\multicolumn{6}{l}{\textit{Graph models without macro features}} \\
GNN-Correlation        & 0.118 & 0.118 & 0.581 & $-$0.128 & 0.606 \\
GNN-Sector             & 0.104 & 0.110 & 0.492 & $-$0.089 & 0.533 \\
GNN-Granger            & 0.096 & 0.112 & 0.412 & $-$0.130 & 0.676 \\
GNN-Ensemble           & 0.106 & 0.112 & 0.498 & $-$0.108 & 0.492 \\
\hline
\multicolumn{6}{l}{\textit{Graph models with macro features}} \\
GNN-Corr.\ + Macro 21d   & 0.137 & 0.105 & 0.828 & $-$0.103 & 0.720 \\
GNN-Corr.\ + Macro 63d   & 0.133 & 0.105 & 0.794 & $-$0.092 & 0.670 \\
GNN-Corr.\ + Macro 126d  & 0.091 & 0.106 & 0.389 & $-$0.126 & 0.884 \\
GNN-Corr.\ + Macro 252d  & 0.117 & 0.101 & 0.671 & $-$0.086 & 0.770 \\
GNN-Sector + Macro        & 0.153 & 0.104 & 0.984 & $-$0.097 & 0.406 \\
GNN-Granger + Macro       & 0.156 & 0.109 & 0.973 & $-$0.101 & 0.499 \\
GNN-Ensemble + Macro      & 0.149 & 0.109 & 0.914 & $-$0.099 & 0.357 \\
\hline
\end{tabular}
\end{table*}

\begin{table*}[t]
\centering
\caption{Inverse-volatility portfolio performance over 103 test weeks with 10 bps transaction cost.}
\label{tab:port_invvol}
\small
\begin{tabular}{lccccc}
\hline
\textbf{Model} & \textbf{Ann.\ Ret.} & \textbf{Ann.\ Vol.} & \textbf{Sharpe} & \textbf{Max DD} & \textbf{Avg.\ Turn.} \\
\hline
\multicolumn{6}{l}{\textit{Baselines}} \\
Equal-weight           & 0.120 & 0.137 & 0.513 & $-$0.171 & 0.000 \\
HAR per-stock          & 0.100 & 0.126 & 0.396 & $-$0.155 & 0.163 \\
HAR pooled             & 0.102 & 0.128 & 0.406 & $-$0.159 & 0.152 \\
LSTM                   & 0.102 & 0.127 & 0.411 & $-$0.158 & 0.076 \\
\hline
\multicolumn{6}{l}{\textit{Graph models without macro features}} \\
GNN-Correlation        & 0.099 & 0.128 & 0.387 & $-$0.159 & 0.079 \\
GNN-Sector             & 0.103 & 0.129 & 0.410 & $-$0.157 & 0.072 \\
GNN-Granger            & 0.105 & 0.129 & 0.423 & $-$0.160 & 0.082 \\
GNN-Ensemble           & 0.103 & 0.128 & 0.417 & $-$0.159 & 0.058 \\
\hline
\multicolumn{6}{l}{\textit{Graph models with macro features}} \\
GNN-Corr.\ + Macro 21d   & 0.108 & 0.128 & 0.451 & $-$0.156 & 0.091 \\
GNN-Corr.\ + Macro 63d   & 0.104 & 0.126 & 0.429 & $-$0.154 & 0.078 \\
GNN-Corr.\ + Macro 126d  & 0.102 & 0.126 & 0.415 & $-$0.154 & 0.113 \\
GNN-Corr.\ + Macro 252d  & 0.103 & 0.125 & 0.425 & $-$0.151 & 0.090 \\
GNN-Sector + Macro        & 0.110 & 0.129 & 0.468 & $-$0.159 & 0.050 \\
GNN-Granger + Macro       & 0.110 & 0.131 & 0.456 & $-$0.164 & 0.053 \\
GNN-Ensemble + Macro      & 0.110 & 0.130 & 0.465 & $-$0.161 & 0.037 \\
\hline
\end{tabular}
\end{table*}

\begin{table*}[t]
\centering
\caption{Volatility-targeted portfolio performance over 103 test weeks with 10 bps transaction cost and 10\% annualized volatility target.}
\label{tab:port_voltarget}
\small
\begin{tabular}{lccccc}
\hline
\textbf{Model} & \textbf{Ann.\ Ret.} & \textbf{Ann.\ Vol.} & \textbf{Sharpe} & \textbf{Max DD} & \textbf{Avg.\ Turn.} \\
\hline
\multicolumn{6}{l}{\textit{Baselines}} \\
HAR per-stock          & 0.038 & 0.053 & $-$0.219 & $-$0.067 & 0.076 \\
HAR pooled             & 0.039 & 0.054 & $-$0.208 & $-$0.069 & 0.073 \\
LSTM                   & 0.044 & 0.055 & $-$0.114 & $-$0.071 & 0.034 \\
\hline
\multicolumn{6}{l}{\textit{Graph models without macro features}} \\
GNN-Correlation        & 0.039 & 0.059 & $-$0.193 & $-$0.076 & 0.052 \\
GNN-Sector             & 0.039 & 0.049 & $-$0.220 & $-$0.061 & 0.028 \\
GNN-Granger            & 0.040 & 0.051 & $-$0.200 & $-$0.066 & 0.037 \\
GNN-Ensemble           & 0.040 & 0.053 & $-$0.196 & $-$0.067 & 0.029 \\
\hline
\multicolumn{6}{l}{\textit{Graph models with macro features}} \\
GNN-Corr.\ + Macro 21d   & 0.045 & 0.060 & $-$0.085 & $-$0.073 & 0.060 \\
GNN-Corr.\ + Macro 63d   & 0.040 & 0.056 & $-$0.172 & $-$0.067 & 0.053 \\
GNN-Corr.\ + Macro 126d  & 0.029 & 0.056 & $-$0.365 & $-$0.067 & 0.083 \\
GNN-Corr.\ + Macro 252d  & 0.030 & 0.056 & $-$0.359 & $-$0.068 & 0.066 \\
GNN-Sector + Macro        & 0.040 & 0.054 & $-$0.189 & $-$0.064 & 0.056 \\
GNN-Granger + Macro       & 0.043 & 0.054 & $-$0.136 & $-$0.069 & 0.036 \\
GNN-Ensemble + Macro      & 0.042 & 0.055 & $-$0.142 & $-$0.068 & 0.037 \\
\hline
\end{tabular}
\end{table*}

\begin{table*}[t]
\centering
\caption{Long-short portfolio performance over 103 test weeks with 10 bps transaction cost.}
\label{tab:port_longshort}
\small
\begin{tabular}{lccccc}
\hline
\textbf{Model} & \textbf{Ann.\ Ret.} & \textbf{Ann.\ Vol.} & \textbf{Sharpe} & \textbf{Max DD} & \textbf{Avg.\ Turn.} \\
\hline
\multicolumn{6}{l}{\textit{Baselines}} \\
HAR per-stock          & $-$0.169 & 0.162 & $-$1.349 & $-$0.359 & 1.330 \\
HAR pooled             & $-$0.192 & 0.153 & $-$1.583 & $-$0.360 & 1.477 \\
LSTM                   & $-$0.157 & 0.165 & $-$1.252 & $-$0.355 & 0.765 \\
\hline
\multicolumn{6}{l}{\textit{Graph models without macro features}} \\
GNN-Correlation        & $-$0.264 & 0.177 & $-$1.779 & $-$0.425 & 0.967 \\
GNN-Sector             & $-$0.170 & 0.175 & $-$1.259 & $-$0.368 & 0.786 \\
GNN-Granger            & $-$0.155 & 0.157 & $-$1.307 & $-$0.344 & 0.983 \\
GNN-Ensemble           & $-$0.185 & 0.182 & $-$1.291 & $-$0.393 & 0.720 \\
\hline
\multicolumn{6}{l}{\textit{Graph models with macro features}} \\
GNN-Corr.\ + Macro 21d   & $-$0.161 & 0.171 & $-$1.237 & $-$0.353 & 1.090 \\
GNN-Corr.\ + Macro 63d   & $-$0.174 & 0.186 & $-$1.206 & $-$0.372 & 0.903 \\
GNN-Corr.\ + Macro 126d  & $-$0.174 & 0.184 & $-$1.217 & $-$0.357 & 0.960 \\
GNN-Corr.\ + Macro 252d  & $-$0.179 & 0.191 & $-$1.202 & $-$0.370 & 0.842 \\
GNN-Sector + Macro       & $-$0.166 & 0.173 & $-$1.250 & $-$0.376 & 0.642 \\
GNN-Granger + Macro      & $-$0.160 & 0.168 & $-$1.251 & $-$0.373 & 0.656 \\
GNN-Ensemble + Macro     & $-$0.149 & 0.175 & $-$1.137 & $-$0.373 & 0.552 \\
\hline
\end{tabular}
\end{table*}

The portfolio results tell four distinct stories, indicating that the same model can succeed on one strategy and fail on another. In minimum variance, graph models with macro features produce the strongest observed results (Table~\ref{tab:port_minvar}). In inverse-volatility weighting, the equal-weight benchmark remains difficult to beat (Table~\ref{tab:port_invvol}). In volatility targeting, the construction produces negative Sharpe ratios across all models after the risk-free adjustment (Table~\ref{tab:port_voltarget}). In long-short, the 2024--2025 regime made the strategy structurally unattractive because many high-volatility stocks also produced strong returns (Table~\ref{tab:port_longshort}). Average weekly turnover shows where aggressive rebalancing creates the most transaction-cost drag.

Minimum variance is the construction where graph models most clearly translate into stronger observed portfolio performance. GNN-Sector + Macro and GNN-Granger + Macro achieve the highest Sharpe ratios, 0.984 and 0.973, compared with 0.635 for HAR per-stock and 0.729 for HAR pooled (Table~\ref{tab:port_minvar}). This shows that the best portfolio model is not necessarily the best point-forecast model. GNN-Corr.\ + Macro 63d has the lowest MSE, but the strongest minimum-variance Sharpe comes from macro-conditioned sector and Granger graphs. The optimizer appears to benefit from the dispersion and cross-sectional structure of those forecasts, rather than average squared error alone.

Inverse-volatility weighting tells a different story. No model beats the equal-weight baseline, whose Sharpe ratio is 0.513 (Table~\ref{tab:port_invvol}). The best model-driven inverse-volatility portfolios are macro GNN variants, led by GNN-Sector + Macro at 0.468, but all remain below equal weight. This suggests that inverse-volatility weighting is more sensitive to calibration and forecast scale than to ranking alone.

Volatility targeting performs poorly across all models, with negative Sharpe ratios throughout Table~\ref{tab:port_voltarget}. GNN-Corr.\ + Macro 21d is the least negative at $-0.085$, followed by LSTM at $-0.114$ and GNN-Granger + Macro at $-0.136$. The strategy reduces realized volatility, but the risk-free-adjusted returns are not strong enough to produce positive Sharpe ratios.

Long-short portfolios are negative across every model in Table~\ref{tab:port_longshort}. The best long-short Sharpe is still negative, at $-1.137$ for GNN-Ensemble + Macro. The construction shorts the highest predicted-volatility stocks, which was costly during 2024--2025 because many high-volatility technology and AI-related stocks also produced strong returns. Turnover was high because the strategy depends on weekly quintile membership: small changes in predicted volatility can move stocks into or out of the long and short legs, forcing frequent rebalancing. This amplified losses through transaction-cost drag, but it was not the root cause. These results show that volatility ranking alone was not a profitable long-short signal in this regime.

\subsection{Synthesis}

Across the full evaluation, no model dominates on every dimension. GNN-Corr.\ + Macro 63d is the best point-forecast model by MSE, GNN-Ensemble + Macro has the strongest mean Rank IC, and GNN-Sector + Macro and GNN-Granger + Macro produce the highest minimum-variance Sharpe ratios. This divergence is the main empirical result: forecast accuracy, ranking quality, and portfolio performance are related but not interchangeable.

The portfolio results also show that forecast improvements do not translate uniformly across construction methods. Minimum-variance optimization is the setting where graph models with macro features show the largest observed improvement over the HAR baselines. In inverse-volatility weighting, volatility targeting, and long-short construction, the gains are weaker or absent. The next section interprets why these differences arise.

\section{Financial Interpretation}

\subsection{Why Volatility Is Predictable At All}

Realized volatility is predictable because it clusters \cite{engle1982}. High-volatility weeks tend to follow high-volatility weeks because shocks do not resolve immediately. When new information hits the market, investors update positions over time, liquidity conditions change, risk limits tighten, and uncertainty remains elevated beyond the initial price move. The result is volatility persistence, where volatility rises quickly after a shock but decays gradually.

This persistence gives HAR a strong baseline advantage. HAR directly models volatility at short, medium, and long horizons, so it captures much of the predictable structure in the target before any graph information is added. The LSTM and GNN models are therefore competing for the remaining signal. Any improvement must come from information HAR does not fully use, such as nonlinear temporal patterns, longer-term dependencies, neighboring stocks, or changing market-wide conditions.

\subsection{When Graph Structure Helps}

Volatility contagion is the mechanism correlation graphs exploit. When a shock hits one stock, it propagates through return co-movement to correlated names, elevating their volatility before it appears in their own history. HAR does not directly anticipate this; a correlation GNN can because it conditions on neighbors' recent behavior directly. Sector and Granger graphs are less effective here because sector membership is fixed annually and Granger edges embed stationarity assumptions that break down when market structure shifts. The rolling correlation graph tracks live co-movement, which is the channel contagion actually travels through.

\subsection{When Graph Structure Hurts}

The same mechanism that makes correlation graphs useful can hurt when market structure changes quickly. The rolling correlation window encodes recent co-movement, not necessarily the current regime. If relationships shift faster than the estimation window adapts, the graph can propagate stale signals from neighbors that are no longer informative. As a result, the model may condition forecasts on relationships that no longer describe the market.

A related risk is oversmoothing. In crisis periods, correlations across stocks spike and the graph becomes dense. The density examples in Table~\ref{tab:graph_density_examples} show why this matters: during the COVID shock, the average stock was connected to 432.7 of 464 possible neighbors, compared with only 42.5 in the sparse 2018 week. A GNN aggregating over many highly correlated neighbors can compress cross-sectional differences in its hidden representations, producing forecasts that are similar across stocks precisely when differentiation matters most for portfolio construction. Monitoring graph density and forecast dispersion during stress periods is a necessary operational check for any deployed graph volatility model. 

\subsection{What Macro Features Contribute}

Macro features provide regime conditioning. VIX and credit spreads signal when the market has entered a stress regime before that stress appears in individual stock volatility histories. This is a different channel from what graph structure provides, since a correlation graph propagates information across stocks within a given week, while macro features tell every stock simultaneously that the broader environment has shifted. Combining the two helps even the weaker graph constructions, as macro conditioning improves GNN-Sector from 0.0336 to 0.0315 and GNN-Granger from 0.0337 to 0.0314, despite leaving their underlying graphs unchanged.

The portfolio results in Table~\ref{tab:port_minvar} make this concrete. Macro conditioning improves not just average forecast error but the distributional properties of predictions the optimizer relies on, which is why GNN-Sector + Macro and GNN-Granger + Macro produce the strongest minimum-variance Sharpe ratios. The forecast-accuracy gains from macro features are notable, but they translate into the largest portfolio improvements where the construction can use the cross-sectional structure those forecasts hold.
 
\subsection{Why the Long-Short Strategy Failed}

In the 2024--2025 test period, many high-volatility technology and AI-related stocks also produced strong returns. Shorting high predicted volatility therefore often meant shorting some of the market's strongest performers. This reflects a limitation of volatility as a standalone return signal in this regime, rather than a failure of forecast accuracy alone.

\subsection{Practical Framework: GNNs vs. HAR}

HAR is competitive with every model in this paper without any cross-stock machinery. When the forecasting task is narrow and the regime is stable, the complexity of a GNN is unlikely to pay off.

GNNs are most justified when the investment problem depends on cross-sectional structure, especially ranking or minimum-variance construction, and when macro features are available to stabilize graph-based forecasts. Their value is not generic model complexity; it comes from giving the portfolio construction useful information that HAR cannot directly represent.

\section{Conclusion}

This paper finds that better volatility forecasts do not automatically lead to better portfolios. Macro-conditioned graph models produce the strongest point-estimate forecasts and the best minimum-variance portfolio results, but their gains are construction-dependent. GNN-Corr.\ + Macro 63d has the lowest MSE, GNN-Ensemble + Macro has the strongest mean Rank IC, and GNN-Sector + Macro has the highest minimum-variance Sharpe. At the same time, equal weight remains stronger than model-driven inverse-volatility portfolios, volatility targeting produces negative Sharpe ratios after the risk-free adjustment, and long-short volatility portfolios perform poorly across every model.

HAR remains a strong default when simplicity, interpretability, and stability matter. GNNs are worth the added complexity when the universe is large, market relationships are informative, macro regime features are available, and the portfolio rule can use cross-sectional forecast structure.

\subsection{Limitations}

Several limitations affect how broadly these results should be interpreted. First, the stock universe is drawn from current S\&P 500 constituents, which introduces survivorship bias by excluding firms that left the index or were delisted before the end of the sample. This likely makes the portfolio backtests cleaner than a fully historical universe would be.

Second, the test period covers only 103 weeks from 2024 through 2025. This period was dominated by a technology- and AI-led market regime in which many high-volatility stocks also produced strong returns. As a result, the poor long-short results should not be interpreted as evidence that volatility ranking can never be profitable, but rather that volatility alone was not a profitable return signal in this regime.

Third, the Granger graph is estimated once and then held fixed. This makes it vulnerable to changes in market structure and may understate the value of causal graph methods that update through time. The rolling correlation graph addresses this issue partially, but it can still lag fast regime shifts if relationships change faster than the estimation window adapts.

Finally, the portfolio results depend on specific implementation choices, including weekly rebalancing, 10 basis point transaction costs, weight constraints, and the selected portfolio constructions. These assumptions are reasonable for comparing models under a controlled framework, but they do not fully capture the frictions, liquidity constraints, and risk controls of a production trading system.
 
\subsection{Future Directions}

Future work should make graph structure more adaptive, especially for Granger edges that are unlikely to remain stable across regimes. GNN architectures should also control oversmoothing through edge dropout, attention, or regime-gated message passing when correlation graphs become dense. Finally, training objectives should better match portfolio use cases by combining point-forecast accuracy with ranking or portfolio-aware losses. Testing the framework across longer samples, earlier crisis periods, and non-U.S. markets would clarify whether the results generalize beyond the 2024--2025 test window.
 
\bibliography{custom}

\end{document}